\documentclass[twocolumn,english,pra,superscriptaddress]{revtex4-2}

\usepackage{color}
\usepackage{amssymb}
\usepackage{amsthm}
\usepackage{graphicx}
\usepackage{epsfig}
\usepackage{subfigure}
\usepackage{amsmath,amssymb,extpfeil,subeqnarray}
\usepackage{enumerate}
\usepackage[colorlinks,citecolor=blue,urlcolor=blue,anchorcolor=blue]{hyperref}
\usepackage{color}
\usepackage{bm}
\usepackage{hyperref}
\hypersetup{
hypertex=true,
colorlinks=true,
linkcolor=blue,
anchorcolor=blue,
citecolor=blue
}
\newcommand{\be}{\begin{equation}}
\newcommand{\ee}{\end{equation}}
\newcommand{\ba}{\begin{eqnarray}}
\newcommand{\ea}{\end{eqnarray}}
\newcommand{\ban}{\begin{eqnarray*}}
\newcommand{\ean}{\end{eqnarray*}}

\begin{document}
	
\title{Quantum Phases of a Dipolar Fermi Gas with Laser-assisted Interwire Tunneling }

\author{Jin-Xin Li}
\affiliation{School of Physics, Henan Normal University, Xinxiang 453007, Henan Province, China}
\author{Xue-Jing Feng}
\affiliation{School of Physics, Henan Normal University, Xinxiang 453007, Henan Province, China}
\author{Ying-Ying Zhang}
\affiliation{School of Physics, Henan Normal University, Xinxiang 453007, Henan Province, China}
\author{Jing-Xue Liu}
\affiliation{School of Physics, Henan Normal University, Xinxiang 453007, Henan Province, China}
\author{Lu Qin}
\affiliation{School of Physics, Henan Normal University, Xinxiang 453007, Henan Province, China}
\author{Zun-Lue Zhu}
\email{zl-zhu@htu.edu.cn}
\affiliation{School of Physics, Henan Normal University, Xinxiang 453007, Henan Province, China}
\author{Xing-Dong Zhao}
\email{phyzhxd@gmail.com}
\affiliation{School of Physics, Henan Normal University, Xinxiang 453007, Henan Province, China}
\author{Liang-Liang Wang}
\email{wangliangliang@westlake.edu.cn}
\affiliation{School of Science, Westlake University, 600 Dunyu Road, Hangzhou 310030, Zhejiang Province, China}
\affiliation{Institute of Natural Sciences, Westlake Institute for Advanced Study, 18 Shilongshan Road, Hangzhou 310024, Zhejiang Province, China}
	
\begin{abstract}
We systematically investigate unconventional superfluid phases of fermionic dipolar particles lying in a double-wire setup with laser-assisted interwire tunneling.
Our numerical simulations, based on the nonlocal Kohn-Sham Bogoliubov-de Gennes equation, reveal the existence of a large Fulde-Ferrell-Larkin-Ovchinnikov (FFLO) region with a stripe phase under an imbalance of particle densities between two wires. 
When the laser-assisted interwire tunneling is present, it induces a transition from the FFLO phase to the topological superfluid phase and the associated Majorana zero modes exhibit an oscillation structure, which is significantly enhanced by the long-range nature of the interwire dipolar interaction.
This distinguishes itself from the results obtained with usual contact interaction and offers new opportunities for manipulating and reshaping Majorana zero modes by adjusting the degree of the nonlocality and the interwire separation.
\par
\end{abstract}

\date{\today}
\maketitle

\section{Introduction}
The search for unconventional superfluid phases in ultracold atomic gases has attracted a great deal of interest, largely due to the existent analogies between neutral superfluids and charged superconductors \cite{science.1255380,science.adg3430, PhysRevLett.116.120403,Qu2013, PhysRevX.9.021025, andp.201700282,PhysRevB.92.035153}.
Of particular relevance are the studies of exotic pairing mechanism in Fermi gases,  e.g., the spatially non-uniform FFLO phase with finite momentum pairing \cite{PhysRev.135.A550, 1965INHOMOGENEOUS,PhysRevB.105.024505,PhysRevB.105.094203,PhysRevB.106.184507,Wan2023,PhysRevA.78.013637, PhysRevA.87.031602,Liao2010-tr, PhysRevA.95.053628,PhysRevA.95.053628,PhysRevB.109.L020504}, as well as the uniform topological superfluidity with non-Abelian Majorana quasiparticles \cite{ PhysRevLett.109.105302, PhysRevLett.113.130404, PhysRevLett.118.207002, PhysRevLett.119.047001,PhysRevA.99.013612,PhysRevA.91.063626, PhysRevA.89.053618,Wang_2020, PhysRevA.107.033326, PhysRevA.88.063601}.
However, a limitation of these gases is that the particle interaction is typically isotropic and extremely short-range contact, making it impossible to exhibit higher-order symmetries.
The production of ultracold dipolar gases with large intrinsic dipole moments promises to change this due to the spatially anisotropic and long-range character of the dipole-dipole interaction \cite{RevModPhys.82.1225, deMiranda2011, PhysRevLett.109.085301, PhysRevA.88.043604, PhysRevLett.114.205302, Li_2017, Kinnunen_2018, Wang_2020J,PhysRevLett.128.223201, PhysRevA.101.033613, New.J.Phys.20.063001,Feng2023}.
Compared to contact interactions, dipole-dipole interactions are neither purely attractive nor purely repulsive.
The attractive pairing channel is mostly of $p_{z}$-like hybridization with higher odd partial wave components.
These features can significantly impact the many-body behaviors of the underlying systems and give rise to unconventional pairing mechanisms \cite {PhysRevLett.105.215302,PhysRevA.83.043602,PhysRevA.96.061602}.

Recently, there has been much interest in quantum dipolar Fermi gases loaded into equidistant wires or layers \cite{Kinnunen_2018,PhysRevA.87.053609,PhysRevA.98.023631, PhysRevLett.108.125301, PhysRevA.98.063610,PhysRevA.101.063624,PRXQuantum.3.030314, PhysRevA.98.043609,PhysRevB.104.045403,s41586-023-05695-4, arXiv2302.07209}.
The single species polarized particles in biwires are connected with each other due to the long-range nature of the dipole-dipole interaction, giving the system a two-component character.
The dipoles in each wire are aligned by an external field and the long-range dipole-dipole interaction can be adjusted by changing the angle between the trap orientation and the polarization direction of the dipoles 
 \cite{Matveeva2011}.
A crossover is also expected by varying the interwire distance, as the system evolves from the weak-coupling normal superfluid regime of largely overlapping Cooper pairs to the strong-coupling BEC regime of composite bosons.
Consequently, these wired structures exhibit remarkable interwire effects such as interwire bound states and non-trivial Cooper pairing of different wire dipolar particles \cite{PhysRevLett.105.215302, PhysRevA.83.043602, PhysRevA.96.061602}.
Additionally, it's well-known that spin-orbit (SO) coupling has been realized by using a pair of counter-propagating Raman lasers \cite{PhysRevA.101.053613, PhysRevA.101.043616, PhysRevLett.117.235304, PhysRevB.97.020501,PhysRevLett.120.060408, PhysRevLett.126.103201, PhysRevLett.109.095301, PhysRevX.6.031022, nphys3672, PhysRevA.105.053312,PhysRevLett.125.260407}.
As well, laser-assisted interwire tunneling, which induce the momentum transition between particles in different wires, can also been achieved by using a pair of counter-propagating Raman lasers \cite{PhysRevA.91.033619,PhysRevA.93.053630,PhysRevA.96.023623, PhysRevA.99.043601, Wang2023,PhysRevResearch.5.L012006,Huang_2016}.
We refer to it as the interwire SO coupling.
To the best of our knowledge, this has not yet been extended to the physically relevant fermionic case when the interwire SO coupling acts together with the long-range dipolar interaction.

Motivated by this, we study here the superfluid phases in a double-wire dipolar fermionic gas with laser-assisted interwire tunneling.
The single-species dipoles are oriented perpendicular to the wires, resulting in repulsive interaction between particles within the same wire.
The effective interwire dipolar interaction is tuned to be attractive and hence generates the interwire pairing 
 \cite{Majorana2006,PhysRevLett.112.086401,PhysRevLett.104.040502,PhysRevB.90.115429}.
The results are based on numerical calculations using the nonlocal Kohn-Sham Bogoliubov-de Gennes equation.
The phase diagram shows a large FFLO phase region with a stripe structure under an imbalance of particle densities of the wires.
Moreover, the system undergoes a quantum phase transition to the topological superfluid phase when the laser-assisted interwire tunneling is present.
The topological nature of the superfluid phase supports Majorana zero modes at the phase boundary, and the long-range nature of dipole interaction induce an oscillatory structure in those zero modes, distinguishing itself from most of the previous results involving only usual contact interaction.
This work thus opens a new direction for manipulating and reshaping Majorana zero modes.

We organize the paper as follows: in Sec.\ref{2}, we introduce the physical model of a dipolar Fermi gas with laser-assisted interwire tunneling and show the scheme. 
In Sec.\ref{3}, the phase diagram is presented according to our analysis of the density distribution and pair profiles.
The presence of the laser-assisted interwire tunneling is found to induce a transition from the FFLO phase to the topological superfluid phase and we then show the energy spectrums and Majorana zero modes for different interwire separations. 
Sec.\ref{4} is devoted to conclusions.

\begin{figure}[t]
\centering
\includegraphics[width=0.49 \textwidth]{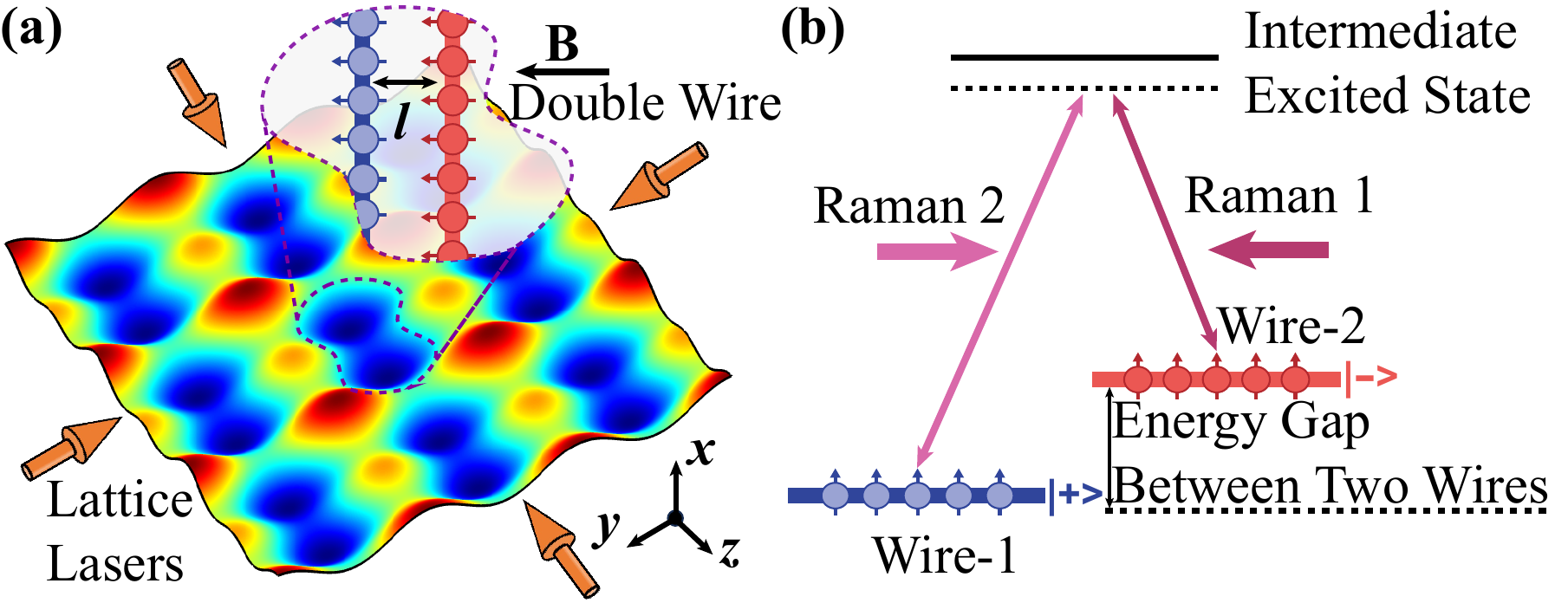} 
\caption{Schematic diagrams for the double-wire system with laser-assisted interwire tunneling. (a) A double-well optical lattice that has two local minima in a unit cell is formed by the interference of two pairs of counter-propagating laser beams. The double-wire system are filled with dipolar Fermi particles oriented perpendicular to the wires, the potential is along \textit{z}-axis, the trapping potential is along the \textit{x}-axis. The blue and red colors signify for the particles in different wire index $\left\{\pm \right\}$. 
(b) The inter-wire tunneling is assisted by a pair of Raman lasers, which induce the momentum transitions between particles in different wires. }
\label{fig-01}
\end{figure}

\begin{figure}[t]
\centering
\includegraphics[width=0.46 \textwidth]{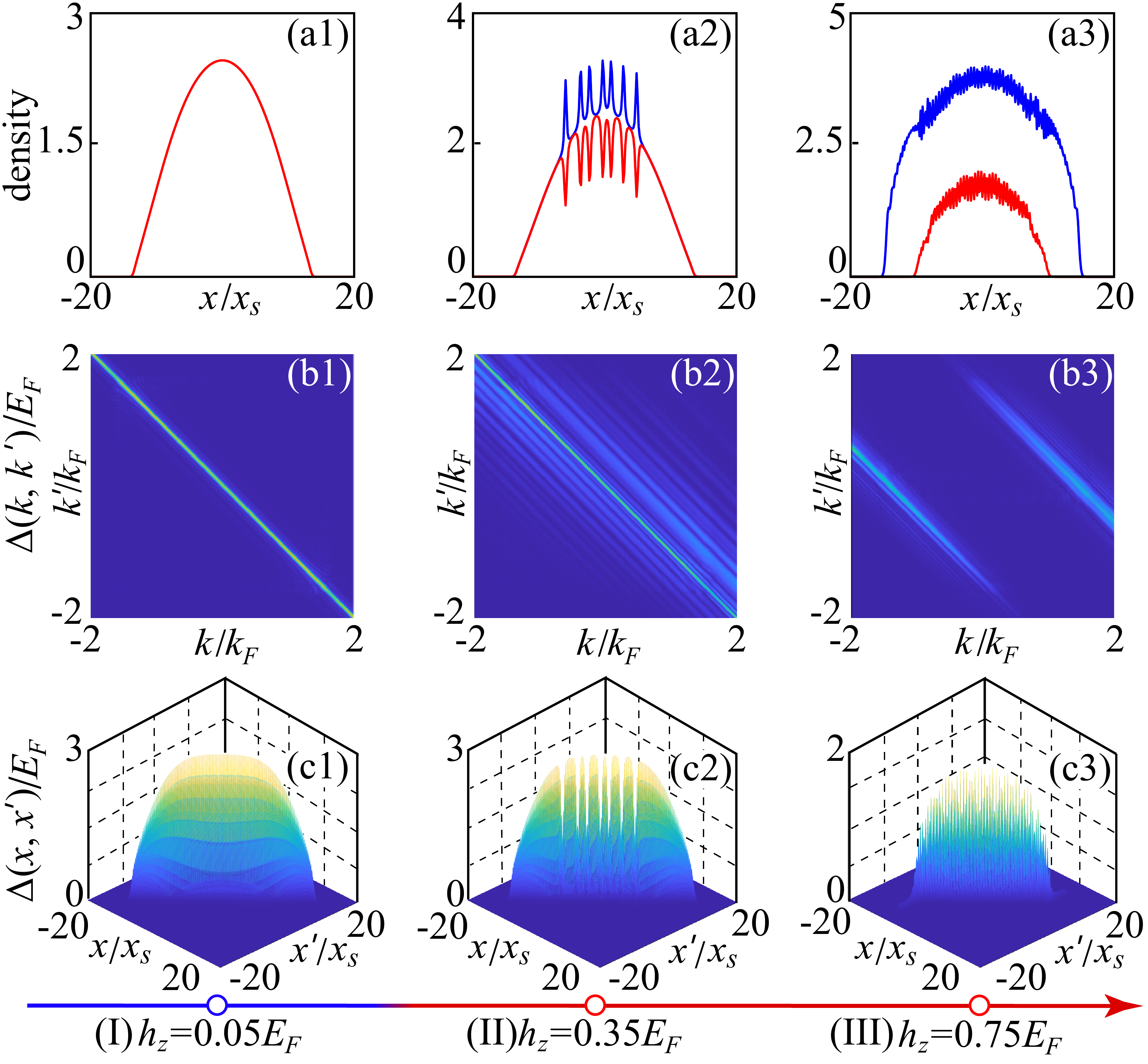} 
\caption{(Upper column) The density profiles of the system with no interwire tunneling for three sets of Zeeman fields (I) 0.25$E_F$, (II) 0.35$E_F$, (III) 0.75$E_F$, where the blue (red) line denotes the density in the wire $+(-)$. The corresponding order parameters in the momentum space $\Delta(k,k')$ (middle column) and real space $\Delta(x,x')$ (down column). Parameters are: $r_{d}=6.25x_s$, $l=0.5x_s$.}
\label{fig-02}
\end{figure}

\section{Hamiltonian and nonlocal KS-BdG formalism}
\label{2}
In this study we consider a dipolar fermionic gas under the laser-assisted tunneling between two wires.
As shown in Fig. \hyperref[fig-01]{1(a)}, the single-component polarized particles with strong dipole-dipole interactions (DDIs) are confined in an asymmetric double-well potential (harmonic) with a distance $l$ along \textit{z}-direction.
Due to the long-range character of the DDIs, the single-species dipole particles in different wires provide a two-component interacting system, which are denoted by the wire index $\left\{\pm \right\}$. 
The particles in each wire are aligned perpendicularly to the \textit{x-y} plane by an magnetic field and the DDIs are well characterized by $V(r_{i},r_{j})=d^2/|r_{i}-r_{j}|^3$,  where $d$ is the magnetic dipolar moment of a particle and $r_{i},r_{j}$ are coordinates.
There is no contact interaction due to Pauli exclusion principle.
The interwire tunneling is restored with two detuned Raman beams, which induce the momentum transfer $\delta k$ between particles in different wires \cite{srep37679,PhysRevA.99.043601, PhysRevA.100.053604, PhysRevLett.117.185301}, as shown in Fig. \hyperref[fig-01]{1(b)}.
Note that the two Raman beams couple different wires but do not change the internal state of the particles.
As a result, the corresponding Hamiltonian of the system is provided as follows:
\begin{equation}
\label{eq1}
\begin{split}
\hat{H}=&\int dx\hat{\rm{\Psi}}^{\dagger}\left\{h_{0}(x)+\lambda k_{x}\sigma_{y} -h_{z}\sigma_{z}\right\}\hat{\rm{\Psi}} \\
+&\sum_{\alpha,\beta}\iint dxdx'\hat{\psi}^{\dagger}_{\alpha}(x)\hat{\psi}^{\dagger}_{\beta}(x')V_{\alpha\beta}(x,x')\hat{\psi}_{\beta}(x')\hat{\psi}_{\alpha}(x),
\end{split}
\end{equation}
where $\hat{\rm{\Psi}}(x)\equiv [\hat{\psi}_{+}(x), \hat{\psi}_{-}(x)]^T$ with the annihilation operator of a fermion at the position $x$ with the wire index $\left\{\pm \right\}$.
$h_{0}(x)=-\frac{\hbar^2}{2m}\partial^2/\partial x^2+V(x)-\mu$, where $V(x)=m\omega^2 x^2/2$ is the harmonic trapping potential with frequency $\omega$, $\mu$ is the chemical potential, and $\sigma_{i\in \{x,y,z\} }$ are the Pauli matrices acting on the wire index. 
The strength of the inter-wire SO coupling $\lambda\equiv \hbar^2\delta k/2m$ can be easily tuned by the Raman lasers, and $h_{z}$ is the effective Zeeman field, adjusted by the particle densities between two wires.
The last term describes the long-range DDIs.
In particular, the intra-wire interaction is purely repulsive, which is given by $V_{++}(x, x')=V_{--}(x, x')=d^2\delta^3/(|x-x'|^2+\delta^2)^{3/2}$ \cite{Elhadj_2019, PhysRevA.93.053627,PhysRevB.96.064513, PhysRevA.94.043638,PhysRevB.98.014513}, where $\delta=\pi^{-1/2}$ is defined as the cut-off parameter to resolve the issue of singularity in $x=x'$.
As a result, it only leads to Fermi renormalization of the density when particles in each wire are in a gas phase, so that below we omit this interaction \cite{PhysRevLett.105.215302, PhysRevA.85.013603}.
The interaction between two wires is more interesting, with a peculiar distribution $V_{+-}(x, x')=d^2(|x-x'|^2-2l^2)/(|x-x'|^2+l^2)^{5/2}$, which is partially attractive and can lead to the inter-wire superfluid pairing \cite{PhysRevLett.105.215302, PhysRevA.83.043602, PhysRevA.96.061602}.
Therefore, we introduce the non-local anomalous density $ \chi(x,x')=\langle\hat{\psi}_{-}(x')\hat{\psi}_{+}(x)\rangle$ \cite{PhysRevLett.60.2430,PhysRevB.72.024545}, which represents the superconducting pairs.
As usual, the Hohenberg-Kohn theorem guarantees a one-to-one mapping between the set of the densities $\lbrace\chi(x, x')\rbrace$ in thermal equilibrium and the set of their conjugate order parameters $\lbrace  \Delta(x,x')\rbrace$, where the nonlocal order parameter as
$\Delta(x,x')=V_{+-}(x,x')\chi(x,x')$, 
with the corresponding effective Hamiltonian can be written as:
$\hat{H}_{\mathrm{eff}}=\int dx \hat{\Psi}^{\dagger}\left\{h_{0}(x)+\lambda k_{x}\sigma_{y}-h_{z}\sigma_{z}\right\}\hat{\Psi}+ \iint dxdx'\Delta^*(x,x') \hat{\psi}_{-}(x')\hat{\psi}_{+}(x)+\mathrm{H.c.}$.
Similar to the standard Hartree-Fock Bogoliubov-de Gennes theory with a local pairing field, the effective Hamiltonian with the nonlocal pairing order parameter  $\Delta(x, x')$ can also be diagonalized by the Nambu transformation:
$\hat{\psi}_{\pm}(x)=\sum_{\eta}\left[ u_{\pm,\eta}(x)\hat{\gamma}_{\eta} + v_{\pm,\eta}^{\ast}(x)\hat{\gamma}_{\eta}^{\dagger}\right]$, where the $\gamma_{\eta}$ is the annihilation operator of quasi-particle in the diagonalized effective Hamiltonian $\hat{H}_{\mathrm{eff}}$ as $\hat{H}_{\mathrm{eff}}=E_{\mathrm{g}}+\sum_{\eta}\epsilon_{\eta}\hat{\gamma}_{\eta}^{\dagger}\hat{\gamma} _{\eta} $, where $E_{\mathrm{g}}$ is the ground state energy and $\epsilon_{\eta}$ is the energy of the $\eta$-th excitation state. 
The resultant nonlocal \textit{Kohn-Sham Bogoliubov-de Gennes} (\textrm{KS-BdG}) equations as  $\int dx' \hat{H}_{\mathrm{KS}}(x,x')\Phi_{\eta}(x')=E_{\eta}\Phi_{\eta}(x)$ \cite{PhysRevLett.60.2430,PhysRevB.72.024545} with $\Phi_{\eta}(x)= [u_{+,\eta}(x), u_{-,\eta}(x), v_{+,\eta}(x), v_{-,\eta}(x)]^{T}$, the nonlocal KS-BdG matrix is:
\begin{widetext}
\begin{equation}
\hat{H}_{\mathrm{KS}}(x,x')=
\begin{bmatrix}
h_{+}(x)\delta(x-x') & -\lambda\frac{\partial}{\partial x}\delta(x-x') & 0 &  \Delta(x,x')\\
\lambda\frac{\partial}{\partial x}\delta(x-x') & h_{-}(x)\delta(x-x') & -\Delta(x',x) &  0\\
0 &  -\Delta^{\ast}(x,x') & -h^{\ast}_{+}(x)\delta(x-x') &  -\lambda\frac{\partial}{\partial x}\delta(x-x')\\
\Delta^{\ast}(x',x) & 0 & \lambda\frac{\partial}{\partial x}\delta(x-x') &  -h^{\ast}_{-}(x)\delta(x-x')\\
\end{bmatrix}
,
\label{(8)}
\end{equation}
\end{widetext}
where $h_{\pm}(x)=h_{0}(x) \mp h_{z}$ and the nonlocal pairing order parameter $\Delta(x,x')$ in Eq.(\ref{(8)}) should be solved self-consistently: $\Delta(x,x')=V_{+-}(x,x')\langle\hat{\psi}_{-}(x')\hat{\psi}_{+}(x)\rangle = (V_{+-}(x,x')/2)\sum_{|E_{\eta}|<E_{c}}[u_{+,\eta}(x)v^{\ast}_{-,\eta}(x')f(E_{\eta})+u_{-,\eta}(x') v^{\ast}_{+,\eta}(x)f(-E_{\eta})]$, and the density in a single wire is given by $n_{\alpha}(x)=\langle\hat{\psi}^{\dagger}_{\alpha}(x)\hat{\psi}_{\alpha}(x)\rangle=(1/2)\sum_{|E_{\eta}|<E_{c}}[ |u_{\alpha,\eta}(x)|^2f(E_{\eta})+|v_{\alpha,\eta}(x)|^2f(-E_{\eta})]$, where $f(E)=1/(1+e^{E/k_{B}T})$ being the quasiparticle Fermi-Dirac distribution with energy $E$ at temperature $T$.

To solve the KS-BdG equation, we convert the equation to a diagonalization problem of a secular matrix by expanding $u_{\pm,\eta}$ and $v_{\pm,\eta}$ on the basis states of the harmonic oscillator.
In our calculation, we set the total number of the atoms to $N = 100$ and the single particle Fermi energy $E_{F}=N\hbar\omega/2$ (neglecting zero point energy), the Fermi momentum $k_{F}=\sqrt{2mE_{F}/\hbar^2}$ and the harmonic oscillator length $x_s=\sqrt{\hbar/(m\omega)}$ are chosen as the units of momentum and length.
The dipolar interaction can be characterized by the effective dipole-dipole distance $r_{d}=md^2/\hbar^2$, which ranges from 10 to $10^4$\r{A} in the laboratory.
In the laser-assisted interwire tunneling case, we choose $\lambda k_{F}=1.5E_{F}$ in most of our calculations.
Up to 400 harmonic oscillator states are involved with an energy cut-off set at $E_c = 15E_{F}$, which is large enough to ensure the accuracy of the calculation for the wavefunction expansion.

\begin{figure}[b]
\centering
\includegraphics[width=0.5 \textwidth]{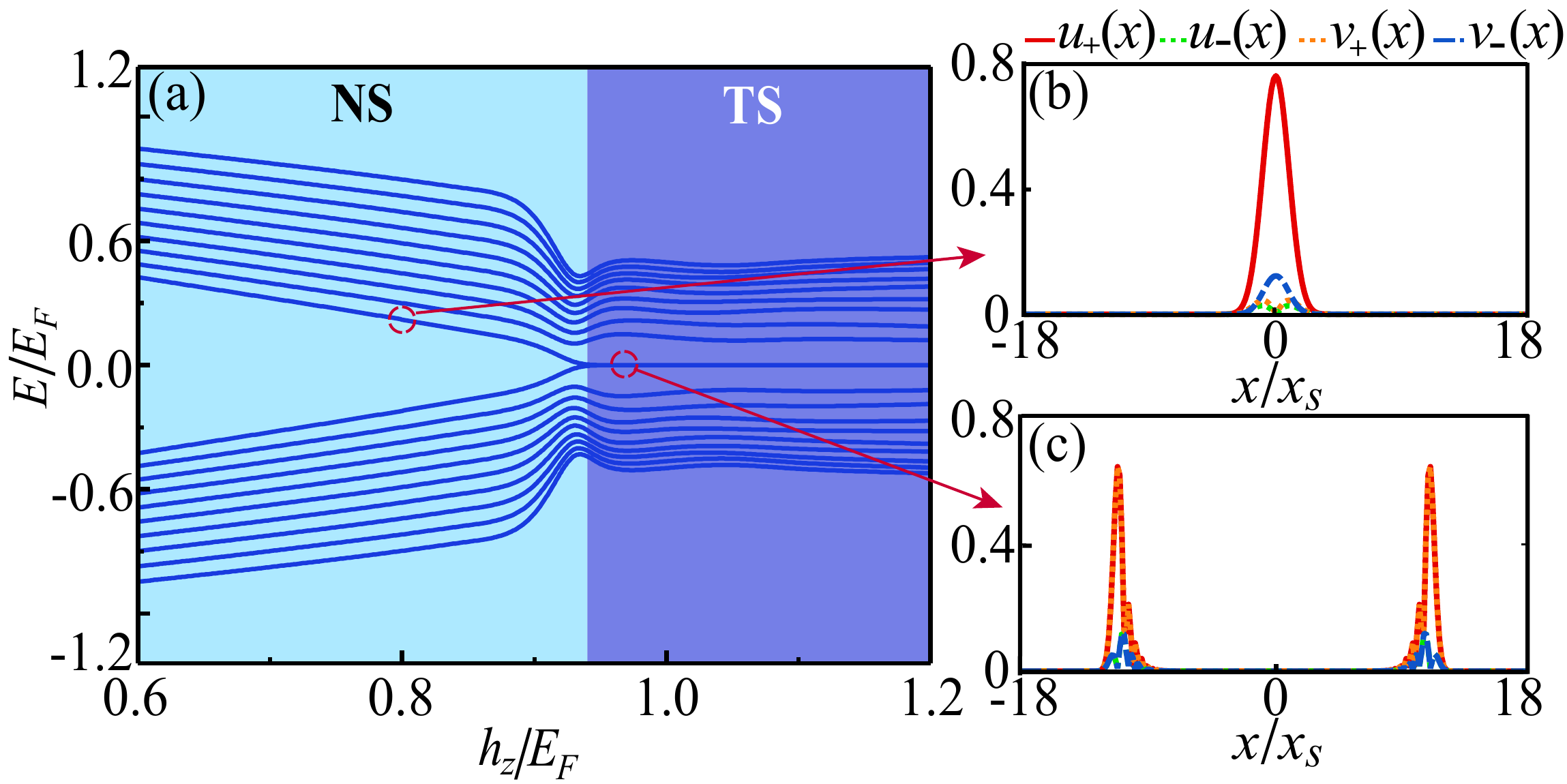} 
\caption{(a) The energy spectrum as a function of the effective Zeeman field for $r_{d}=6.25x_s$, $l=0.5x_s$, $\lambda k_F=1.0E_F$. The energy gap get close at Zeeman field $h_z=0.94E_F$ with a phase transition from NS state to the topological superfluid (TS) phase. The wavefunctions of the lowest energy state are shown for these two phases at Zeeman fields (b)$0.8E_F$ and (c)$0.99E_F$ respectively.}\label{fig-03}
\end{figure}

\begin{figure}[t]
\centering
\includegraphics[width=0.46\textwidth]{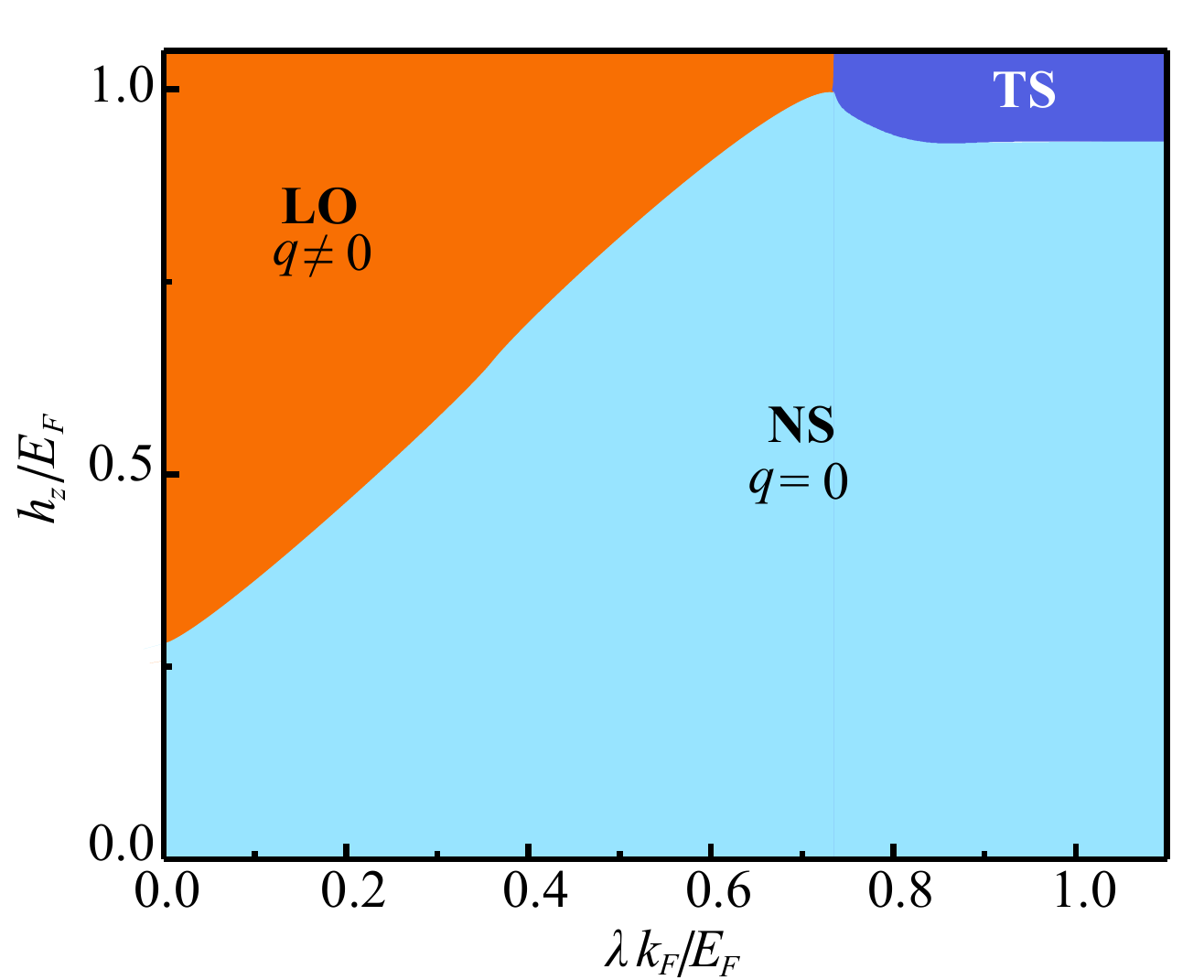} 
\caption{The phase diagram in terms of $\lambda k_{F}/E_{F}$ and $h_{z}/E_{F}$ with $r_{d}=6.25x_s$ and $l=0.5x_s$. LO phase: FFLO superfluid with stripe phase in the orange region; NS: normal superfluid in the green region; TS: topological superfluid with the appearance of MZMs in the blue region.}
\label{fig-04}
\end{figure}
\section{Results}
\label{3}

We first start with no interwire tunneling.
In this case we find that the particles located in the different wires are only coupled to each other by long-range dipolar interactions.
Fig. \ref{fig-02} displays the nonlocal order parameters both in real and momentum space and corresponding density profiles with different effective Zeeman fields.
As anticipated, the ground state is in a normal superfluid with zero momentum pairing under weak Zeeman fields.
For a given dipole-dipole distance, the FFLO phase emerges over a critical Zeeman field value and our numerical calculations show that this is a tripe phase.
Non-zero momentum Cooper pairing yields a spatially dependent superfluid order parameter along the diagonal directions, as shown in Fig. \hyperref[fig-02]{2(c2-c3)}.
This is a key feature of the so-called LO phase.
Thus the LO part of the nonlocal superfluid order parameter can be characterized by $\Delta(R)=\sum_{i}|\Delta_{Q_{i}}(R)| e^{iQ_{i} \cdot R}$, where $Q_{i}$ is the relative momentum of the two paired particles, and $R$ is their center-of-mass (CM) position of the Cooper pairing.
As expected, two symmetric off-diagonal peaks appear in the pair momentum distribution $\Delta(k,k')$ with $Q_{1}=-Q_{2}\neq 0$, which serves as an evidence for the stripe FFLO state with long-range interaction.
In Fig. \hyperref[fig-02]{2(b2-b3)}, we observe as the Zeeman field increases, the relative momentum of the Cooper pairing increases, but the amplitude of the superfluid order parameter decreases.
For a sufficiently large field, the pairing will eventually be destroyed and the ground state will become the normal state.  
The behaviors of the density profiles are similar to the nonlocal order parameters.
As one can see In Fig. \hyperref[fig-02]{2(c2)}, the wings of the density profile are not polarized and the core of the density is filled with an oscillating structure.
Since the no-polarized edge is in the trivial superfluid phase, the FFLO state we studied is a mixed phase.
However, without making confusion, we still call it the FFLO state.

Let us now investigate the system with laser-assisted interwire tunneling.
The state between two wires are mixed via the laser-assisted tunneling and the energy gap $E_{\mathrm{gap}}=2h_{z}$ appears in the single-particle spectrum.
With the long-range dipole-dipole interaction, topological transition might appear and Majorana zero modes may be observed.
As a necessary process to change the topology of the system, we first observe the closing of the energy gap.
In Fig. \hyperref[fig-03]{3(a)}, we present the behavior of the energy spectrum while increasing the Zeeman field.
It is evident that the excitation gap vanishes across the critical point, indicating the topological phase transition.
Next we explore the structure of the zero-energy states.
Because of the intrinsic particle-hole symmetry of the KS-BdG formalism, one can immediately get the zero-energy quasiparticle states satisfy $\gamma_{0}=\gamma_{0}^{\dagger}$, thus these modes can be regarded as Majorana quasiparticles.
The ground state is degenerate: $\gamma_{0}|GS\rangle=0$ and $|GS\rangle=\gamma_{0}^{\dagger}|GS\rangle$.
These two degenerate states can be used as a qubit for quantum information processing.
The corresponding wave functions for these zero-energy states should satisfy either $u_{\sigma}(x)=v^{\ast}_{\sigma}(x)$ or $u_{\sigma}(x)=-v^{\ast}_{\sigma}(x)$.
The wave functions of the zero-energy modes are shown in Fig. \hyperref[fig-03]{3(c)}, we can see that the wave functions readily satisfy the requirement of symmetry or anti-symmetry and intrinsically nonlocal, with weight at two spatially separated points.

Based on the above analysis, we address the ground-state phase diagram with respect to $\lambda$ and $h_{z}$, as shown in Fig. \ref{fig-04}.
Three phases are obtained: normal-superfluid phase (NS) with ($\Delta\neq0$, $Q_{i}=0$), stripe FFLO superfluid phase (LO) with ($\Delta\neq0$, $Q_{i}\neq0$) and topological superfluid phase (TS) with Majorana zero modes (MZMs). 
Similar to that in the case of contact interactions, the transition from the normal superfluid to FFLO state is of the first order.
Remarkably, the FFLO region significantly depends on the strength of the interwire tunneling and survives in a large regime due to the long-range character of the dipolar interaction.
A key point is that the the long-range feature of the dipolar interaction can be modified by the wire separation $l$.
The potential $V_{+-}(x)$ is attractive with $x<\sqrt{2} l$, while the inter-wire distance is increased, the effective range of the nonlocal superfluid pairing is shifted and thus, the corresponding topology is changed.

\begin{figure}[t]
\centering
\includegraphics[width=0.489 \textwidth]{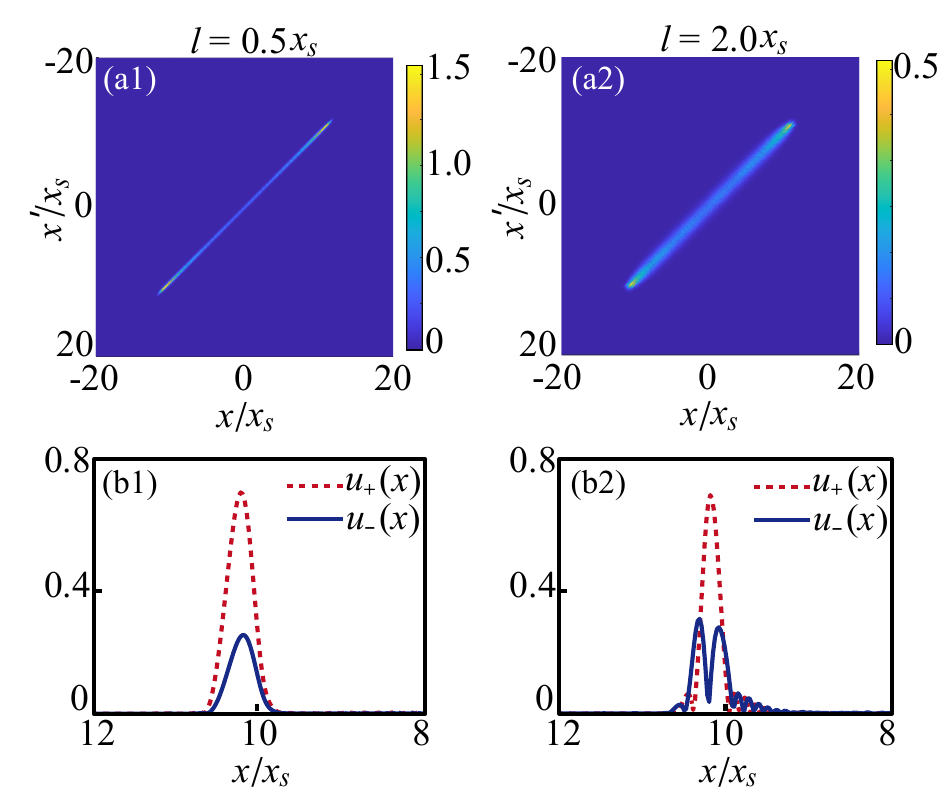} 
\caption{The distributions of the nonlocal order parameters (a) and the corresponding wavefunctions of the MZMs (b), for different interwire separations as shown in the figure. Solid blue line for $u_{-}$ and dashed red one for $u_{+}$. Here  $\lambda k_{F}=1.5E_{F}$, $r_{d}=6.25x_s$, $h_{z}=1.0E_{F}$.}
\label{fig-05}
\end{figure}

\begin{figure}[t]
\centering
\includegraphics[width=0.45\textwidth]{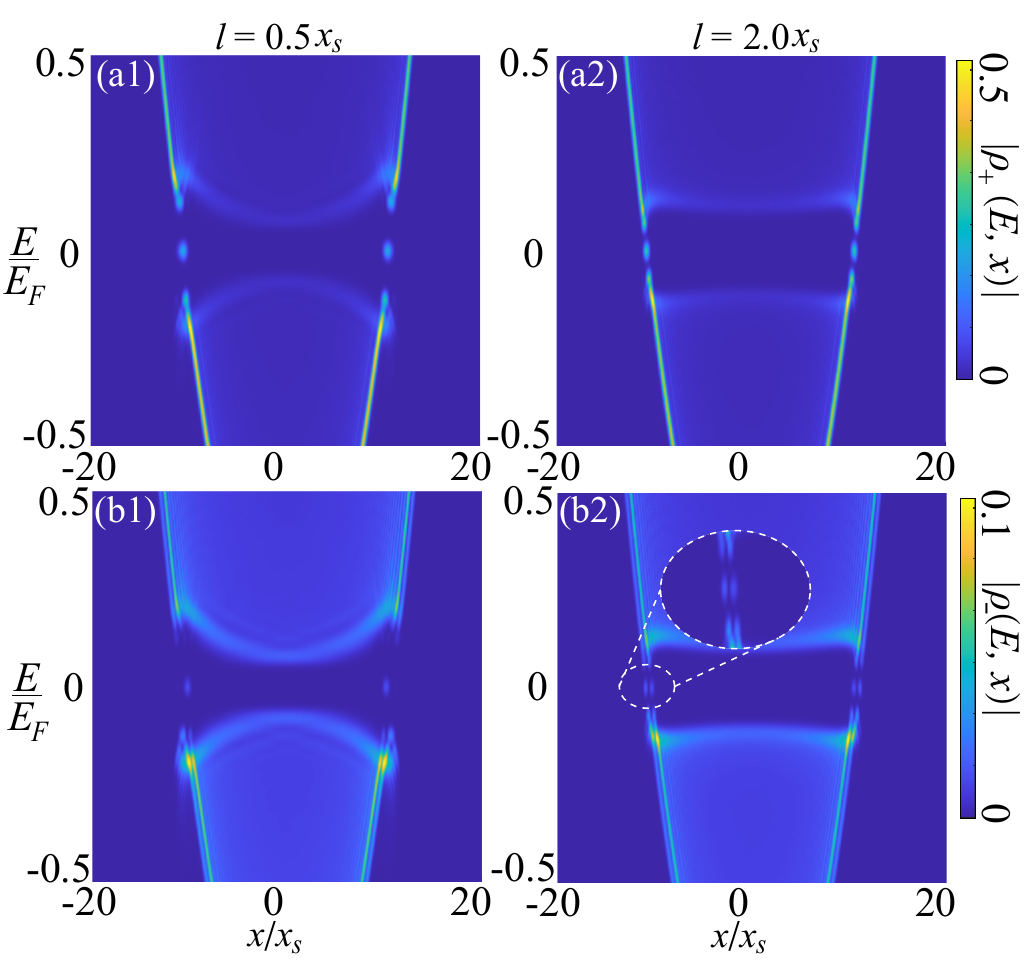} 
\caption{Linear contour plot for the local density of states of particles in different wires $\rho_{\pm}(x)$ for the case in Fig. \hyperref[fig-05]{5}. The signals of Majorana zero modes are well isolated in the energy and spatial domain, which have been zoomed in to get more details in the white circle area.}
\label{fig-06}
\end{figure}

In Fig. \hyperref[fig-05]{5}, we examine the role of the interwire distance in the behavior of the MZMs.
First, it is worth observing from the Fig. \hyperref[fig-05]{5(a1-a2)} that the non-localization of the pairing order parameter $\Delta(x,x')$ is enhanced with the wire separation.
For both $l=0.5x_s$ and $l=2x_s$, two MZMs emerge at the phase interfaces $x \approx \pm10x_s$ and well isolated from other quasiparticles.
Note that the overlap between the wave functions of two MZMs lead to finite but exponentially small energy splitting: $E_{0}\approx1 \times10^{-4}E_{F}$.
For a better observation of the structure of MZMs, we just illustrate the wave functions of MZM at $x \approx 10x_{s}$.
Similar to that in the case of contact interaction, the wave functions of MZMs under small wire separation (Fig. \hyperref[fig-05]{5(b1)}) decreases smoothly with steep slope.
Be contrast, there is an oscillation structure of the MZMs with larger wire separation, which is considerably distinct for the $u_{-}(x)$.
As shown in Fig. \hyperref[fig-05]{5(b2)}, two obvious peaks in the distribution of $u_{-}(x)$ should be detected in the local density of states, which can be measured by momentum resolved radio-frequency spectroscopy.
The local density of states for particles in different wires $\rho_{\alpha}(x,E)$ is defined as $\rho_{\alpha}(x,E)=1/2\sum_{\eta}|u_{\alpha,\eta}(x)|^2\delta(E-E_{\eta})+|v_{\alpha,\eta}(x)|^2\delta(E+E_{\eta})$.
In Fig. \ref{fig-06}, we display corresponding $\rho_{\alpha}(x,E)$ and the contribution from MZMs are visible and well separated from other quasi particle contributions.
Clearly, two peaks of $u_{-}(x)$ appear at the interface with larger wire separation, which distinguishes itself from most of the previous results.

\section{CONCLUDING REMARKS}
\label{4}

In this work, we utilized the KS BdG formulation to study possible FFLO and topological phases in a double-wire system of dipolar Fermi particles with the laser-assisted tunneling between two wires.
The particles are oriented perpendicular to the plane, and the superfluid pairing is provided by the partially attractive long-range interaction between dipoles belonging to different wires.
In the absence of laser-assisted interwire tunneling, a large FFLO region with a stripe phase is found under an imbalance of particle densities of the wires.
A transition from the FFLO phase to the topological superfluid phase occurs when the inter-wire laser-assisted tunneling is turned on.
It is noteworthy that the associated Majorana zero modes exhibit an evident oscillation of wave functions and change as the interwire distance is varied.
Our work demonstrates the importance of the long-range character of the DDI, which enhances the oscillation structure of Majorana zero modes and supports the FFLO regime.
With the techniques of dipolar Fermi particles with strong dipolar interaction by Dy \cite{arXiv2302.07209} and laser-assisted interwire tunneling by Raman laser \cite{PhysRevResearch.5.L012006}, our scheme is within the reach of current experiments.
We hope that the insights obtained in this work may offer intriguing perspectives for exploring new physics phenomena in dipolar systems.

\section*{Acknowledgements}
We thank Wenjun Shao for the early contribution to this project and are grateful to Lin Wen and K. A. Yasir for valuable discussions. This work is supported by National Natural Science Foundation of China (12234012).  The numerical calculations in this paper have been done on the supercomputing system in the Information Technology Center of Westlake University.

\end{document}